\documentclass[graybox]{svmult}

\usepackage{type1cm}        
\usepackage{makeidx}         
\usepackage{graphicx}        
\usepackage{multicol}        
\usepackage[bottom]{footmisc}

\usepackage{newtxtext}       %
\usepackage{newtxmath}       

\DeclareMathOperator{\num}{num}
\DeclareMathOperator{\den}{den}
\DeclareMathOperator{\re}{Re}
\DeclareMathOperator{\im}{Im}
\DeclareMathOperator{\sgn}{sgn}

\newcommand{\HA}{\mathop{\mathrm{H}}\nolimits}
\newcommand{\M}[2]{\mathop{\mathbf{M}}\!\left(#1\right)\left(#2\right)}
\newcommand{\MMA}{\mbox{\textsc{Mathematica}}}
\newcommand{\Maple}{\mbox{\textsc{Maple}}}
\newcommand{\Singular}{\mbox{\textsc{Singular}}}

\usepackage{hyperref}

\begin{document}

\title*{Nested Integrals and Rationalizing Transformations}
\author{Clemens G. Raab}
\institute{Clemens G. Raab \at Johannes Kepler Universit\"at Linz, Inst. f. Algebra, Altenberger Stra{\ss}e 69, 4040 Linz, Austria \email{clemensr@algebra.uni-linz.ac.at}}
%
%
\maketitle

\abstract{A brief overview of some computer algebra methods for computations with nested integrals is given. The focus is on nested integrals over integrands involving square roots. Rewrite rules for conversion to and from associated nested sums are discussed. We also include a short discussion comparing the holonomic systems approach and the differential field approach. For simplification to rational integrands, we give a comprehensive list of univariate rationalizing transformations, including transformations tuned to map the interval $[0,1]$ bijectively to itself.}

\section{Introduction}

By nested integrals, we mean multiple integrals whose integrand is a product of individual integrands depending only on one integration variable each and where each integration variable occurs as an integration bound for the next inner integral. Commonly, they use the origin as their base point and have the form
\begin{equation}\label{eq:NestedIntegral0}
 \int_0^xf_1(t_1)\int_0^{t_1}f_2(t_2)\dots\int_0^{t_{k-1}}f_k(t_k)\,dt_k\dots dt_1,
\end{equation}
where each $f_i$ is allowed to depend only on $t_i$ and on external parameters, but not on any other integration variable $t_j$. Note that here and in all that follows, the possible dependence on external parameters is not denoted explicitly. Also conventions with base points other than the origin are possible, e.g.
\begin{equation}\label{eq:NestedIntegral1}
 \int_x^1f_1(t_1)\int_{t_1}^1f_2(t_2)\dots\int_{t_{k-1}}^1f_k(t_k)\,dt_k\dots dt_1.
\end{equation}

Choosing the integrands from certain classes of functions, the nested integrals give rise to various classes of functions. Already Kummer \cite{Kummer} considered nested integrals over rational functions and determined that it is sufficient to consider integrands that are reciprocals of linear polynomials in the integration variable. These integrals are often referred to as \emph{hyperlogarithms}. They occurred also in the works of Poincar\'{e}, Lappo-Danilevski, and many others, see e.g.\ \cite{Poincare,LappoDanilevski,Wechsung,Goncharov}. In the context of QFT, special choices of rational integrands have been used frequently, giving \emph{harmonic polylogarithms} \cite{RemiddiVermaseren}, \emph{cyclotomic harmonic polylogarithms} \cite{Cyclotomic}, and \emph{generalized harmonic polylogarithms} \cite{HPL}. Generalizing the notation of Ref.~\cite{RemiddiVermaseren}, integrands for generalized harmonic polylogarithms $\HA\nolimits_{a_1,\dots,a_k}(x)$ are given by
\begin{equation}\label{eq:IntegrandRat}
 f_a(x) := \frac{c_a}{x-a}
\end{equation}
with $c_a:=\sgn(-a+0)$ such that $f_a(x)>0$ for $x>0$ close to zero, which is the base point of these integrals following \eqref{eq:NestedIntegral0}. For integrals of the form \eqref{eq:NestedIntegral1}, the choice $c_a:=\sgn(1-a-0)$ is preferred in order to have $f_a(x)>0$ for $x<1$ close to one.

Going beyond rational integrands, one can also allow more general functions by requiring only the square of the integrand to be a rational function. A minimal set of integrands needed to express these integrals (and hence also those over any algebraic functions expressible by non-nested square roots) was determined in Ref.~\cite{JMP2014} based on work by Hermite \cite{HermiteSqrt}, which consists of integrands \eqref{eq:IntegrandRat} and the following ones.
\begin{align}
 f_{\{a_1,\dots,a_k\}}(x)&:=f_{a_1}(x)^{1/2}\dots f_{a_k}(x)^{1/2}\quad\quad\text{for }k\ge2\label{eq:IntegrandSqrt1}\\
 f_{(a,\{a_1,\dots,a_k\})}(x)&:=f_a(x)f_{\{a_1,\dots,a_k\}}(x)\quad\quad\text{for }k\ge1, a\not\in\{a_1,\dots,a_k\}\label{eq:IntegrandSqrt2}\\
 f_{(\{a_1,\dots,a_k\},j)}(x)&:=x^jf_{\{a_1,\dots,a_k\}}(x) \quad\quad\text{for }j\in\{1,\dots,k-2\}\label{eq:IntegrandSqrt3}
\end{align}
In the context of QFT, certain explicit non-rational integrands of this type occurred already in Ref.~\cite{AgliettiBonciani}, for instance, and continue to arise more often in computations in recent years, see e.g.\ \cite{NPB2014,NPB2019a,NPB2019b,NPB2020}.

More generally, one can use integrands that are \emph{hyperexponential} functions, i.e.\ their logarithmic derivative is a rational function. Rational functions and square roots of rational functions mentioned so far are hyperexponential too. Nested integrals over hyperexponential functions give rise to the \emph{d'Alembertian functions} \cite{AbramovPetkovsek}.

In the context of QFT, also nested integrals with other types of integrands beyond hyperexponential functions arise as well. For instance, integrands appear that are solutions of second-order differential equations represented in terms of complete elliptic integrals or ${}_2F_1$-functions, see e.g.\ \cite{JMP2018}.

It is well known that, for reasonably regular integrands, nested integrals satisfy the shuffle relations \cite{Ree}. These relations apply, for instance, if integrands have at most simple poles. In general, however, additional terms arise in the shuffle relations, which are given by nested integrals of lower depth. An algebraic theory covering that is worked out by the author and Georg Regensburger in Ref.~\cite{FTC}.

For a given class of integrands, if one chooses a minimal set of integrands among them such that still all nested integrals of that type can be expressed, then one can deduce that nested integrals over that set of integrands do not satisfy any additional algebraic relation beyond the shuffle relations. This in turn allows to compute canonical forms of quantities that are expressed polynomially in terms of nested integrals over the given class of integrands. For rational integrands, for example, the set of all integrands of the form \eqref{eq:IntegrandRat}, for $a \in \mathbb{C}$, has this property. Hence, over the rational functions, all algebraic relations of generalized harmonic polylogarithms are given by the shuffle relations, cf.\ also Ref.~\cite{DeneufchatelEtAl}.
The analogous statement for nested integrals over functions whose square is rational holds when one selects the set of integrands given by Eqs. \eqref{eq:IntegrandRat} through \eqref{eq:IntegrandSqrt3} and branch cuts are ignored. A minimal set of hyperexponential integrands needed to express all d'Alembertian functions as well as a corresponding canonical form was worked out by the author \cite{RaabMalaga}.

For the rest of this chapter, we focus on integrands whose square is rational. In Section~\ref{sec:sums2integrals}, by briefly looking at how nested integrals may arise from nested sums, we highlight computational methods introduced in Ref.~\cite{JMP2014} relying on special identities that have been constructed to be used as rewrite rules. These rewrite rules allow to do certain computations with nested integrals and nested sums more directly than via e.g.\ constructing and solving differential equations as it is done in more general methods. We briefly compare two computer algebra approaches to the construction of such differential equations. Section~\ref{sec:transformations} is devoted to expressing nested integrals involving square roots in terms of hyperlogarithms by suitable change of variables.

Many of the formulae and algorithms discussed below are implemented in the package \texttt{HarmonicSums} \cite{AblingerDipl,AblingerPhD,HarmonicSums}.

\section{Obtaining nested integrals from nested sums}
\label{sec:sums2integrals}

In analogy to nested integrals, nested sums are multiple sums of a summand that is the product of summands that depend only on one summation variable and where each summation variable occurs as one bound of the summation range of the next inner sum. For example, they may take the general form
\[
 \sum_{i_1=1}^nf_1(i_1)\sum_{i_2=1}^{i_1}f_2(i_2)\dots\sum_{i_k=1}^{i_{k-1}}f_k(i_k).
\]

Passing from sequences indexed by a discrete variable $n$ to functions depending on a continuous variable $x$ can work in two main ways. On the one hand, we can view the sequence given as the sequence of Taylor coefficients of a function, the \emph{generating function} of the sequence. On the other hand, we can aim at an integral representation of the sequence, i.e.\ a definite integral over an integrand depending on a parameter $n$ that reproduces the entries of the sequence.
Often, such integral representations are based on integral transforms. Below, we utilize the Mellin transform, modified to take the following form.
\begin{equation}\label{eq:Mellintransform}
 \M{f(x)}{n} = \int_0^1x^nf(x)\,dx
\end{equation}

If the summands $f_j(n)$ are \emph{P-finite} (also called \emph{P-recursive}) sequences, i.e.\ sequences that satisfy a linear recurrence with polynomial coefficients, then so are the corresponding nested sums. Closely related is the concept of \emph{holonomic} sequences, which is equivalent in case of univariate sequences. In particular, for summands satisfying first-order recurrences the nested sums are \emph{d'Alembertian} sequences \cite{AbramovPetkovsek}, like the harmonic sums \cite{Vermaseren}, generalized harmonic sums (S-sums) \cite{MochUwerWeinzierl}, or nested (inverse) binomial sums \cite{JMP2014}, for example. This allows general strategies and algorithms for P-finite/holonomic sequences to be applied to nested sums as well.

In the following, however, our focus lies on approaches and algorithms that exploit the nested structure of the sums. Similarly, also for integrals, there are specialized methods that are able to exploit the structure of nested integrals, in addition to the general algorithms that do not. Such dedicated approaches not only reduce the computational burden, but also enable general theoretical statements to be proven that would be very hard to obtain otherwise, see also \cite{JMP2014}.

\subsection{Generating functions}

Generating functions are defined by infinite sums of the form
\begin{equation}
 F(x) = \sum_{n=0}^\infty f_nx^n.
\end{equation}
It is well known that any linear recurrence with polynomial coefficients for the sequence $(f_n)$ can be translated into a linear differential equation with polynomial coefficients for the function $F(x)$. This provides a general strategy to compute the function $F(x)$ by constructing and solving a differential equation, starting form a recurrence for the sequence $(f_n)$, independent of the explicit form of that sequence. In practice, however, constructing and solving differential equations can be avoided altogether in many cases by exploiting the syntactic presentation of the sequence. 

This can be done by utilizing general properties of generating functions that can be interpreted as rewrite rules. The aim of such rewrite rules is to express $F(x)$ in terms of other generating functions of sequences that are simpler. For instance, a well-known identity for generating functions is 
\begin{equation}\label{eq:rule701}
 \sum_{n=1}^\infty x^n\frac{f_n}{n} = \int_0^x\frac{1}{t}\sum_{n=1}^\infty t^nf_n\,dt,
\end{equation}
which holds for arbitrary sequences $(f_n)$. It reduces computing the generating function of a sequence $(\frac{f_n}{n})$ to computing the generating function of $(f_n)$. In general, one is interested in identities that allow to simplify generating functions of the forms $\sum_{n=0}^\infty x^ng_nf_n$ and $\sum_{n=0}^\infty x^ng_n\sum_{i=0}^nf_i$ for concrete $g$ but arbitrary $f$. In Eq.~\eqref{eq:rule701}, we have $g$ such that $g_0=0$ and $g_n=\frac{1}{n}$ for $n\ge1$.

Eqs.~(7.1) through (7.11) in Ref.~\cite{JMP2014} give rewrite rules for evaluating generating functions involving sums. Among them, the rules
\begin{align}
 \sum_{n=1}^\infty x^n\binom{2n}{n}\sum_{i=1}^nf_i &= \frac{1}{4\sqrt{\frac{1}{4}-x}}\int_0^x\frac{1}{t\sqrt{\frac{1}{4}-t}}\sum_{n=1}^\infty t^nn\binom{2n}{n}f_n\,dt\label{eq:rule705}\\
  \sum_{n=1}^\infty \frac{x^n}{n\binom{2n}{n}}\sum_{i=1}^nf_i &= \frac{\sqrt{x}}{\sqrt{4-x}}\int_0^x\frac{1}{\sqrt{t}\sqrt{4-t}}\sum_{n=0}^\infty \frac{t^n}{\binom{2n}{n}}f_{n+1}\,dt\label{eq:rule706}\\
   &= \sum_{n=1}^\infty \frac{x^n}{n\binom{2n}{n}}f_n+\frac{\sqrt{x}}{\sqrt{4-x}}\int_0^x\frac{1}{\sqrt{t}\sqrt{4-t}}\sum_{n=1}^\infty \frac{t^n}{\binom{2n}{n}}f_n\,dt\label{eq:rule707}
\end{align}
as well as
\begin{equation}\label{eq:rule708}
 \sum_{n=1}^\infty\frac{x^n}{(2n+1)\binom{2n}{n}}\sum_{i=1}^nf_i = \frac{2}{\sqrt{x}\sqrt{4-x}}\int_0^x\frac{1}{\sqrt{t}\sqrt{4-t}}\sum_{n=1}^\infty\frac{t^n}{\binom{2n}{n}}f_n\,dt
\end{equation}
involve also the central binomial coefficient.
We illustrate the use of such rewrite rules by the following small example.

\begin{example}
Consider the generating function given by
\[
 \sum_{n=1}^\infty x^n\frac{1}{n^2\binom{2n}{n}}\sum\limits_{i=1}^n\frac{1}{i}.
\]
In order to apply one of the rules involving the binomial coefficient, we first need to use Eq.~\eqref{eq:rule701}. Applying that rule to $f_n=\frac{1}{n\binom{2n}{n}}\sum\limits_{i=1}^n\frac{1}{i}$, we obtain
\[
 \sum_{n=1}^\infty x^n\frac{1}{n^2\binom{2n}{n}}\sum\limits_{i=1}^n\frac{1}{i} = \int_0^x\frac{1}{t}\sum_{n=1}^\infty t^n\frac{1}{n\binom{2n}{n}}\sum\limits_{i=1}^n\frac{1}{i}\,dt.
\]
Proceeding with the new generating function inside the integrand, we apply Eq.~\eqref{eq:rule707} to $f_n=\frac{1}{n}$ in order to obtain
\[
 \sum_{n=1}^\infty x^n\frac{1}{n\binom{2n}{n}}\sum\limits_{i=1}^n\frac{1}{i} = \sum_{n=1}^\infty \frac{x^n}{n^2\binom{2n}{n}}+\frac{\sqrt{x}}{\sqrt{4-x}}\int_0^x\frac{\sum\limits_{n=1}^\infty \frac{t^n}{n\binom{2n}{n}}}{\sqrt{t}\sqrt{4-t}}\,dt.
\]
To treat the first term on the right hand side, we apply Eq.~\eqref{eq:rule701} with $f_n=\frac{1}{n\binom{2n}{n}}$ yielding
\[
 \sum_{n=1}^\infty \frac{x^n}{n^2\binom{2n}{n}} = \int_0^x\frac{1}{t}\sum_{n=1}^\infty \frac{t^n}{n\binom{2n}{n}}\,dt.
\]
It remains to evaluate the last generating function by instantiating Eq.~\eqref{eq:rule706} with the Kronecker delta $f_n=\delta_{1,n}$.
\[
 \sum_{n=1}^\infty \frac{x^n}{n\binom{2n}{n}}\sum_{i=1}^n\delta_{1,i} = \frac{\sqrt{x}}{\sqrt{4-x}}\int_0^x\frac{\sum\limits_{n=0}^\infty \frac{t^n}{\binom{2n}{n}}\delta_{1,n+1}\,dt}{\sqrt{t}\sqrt{4-t}} = \frac{\sqrt{x}}{\sqrt{4-x}}\HA_{\{0,4\}}(x)
\]
Recall that the integrand $f_{\{0,4\}}$ is defined by Eqs. \eqref{eq:IntegrandRat} and \eqref{eq:IntegrandSqrt1} so that one could also write $\HA_{\{0,4\}}(x)=\arccos(1-\frac{x}{2})$ explicitly. Altogether, we obtained the generating function as a sum of two nested integrals performing hardly any computation and without constructing any differential equation.
\[
 \sum_{n=1}^\infty x^n\frac{1}{n^2\binom{2n}{n}}\sum\limits_{i=1}^n\frac{1}{i} = \HA_{0,\{0,4\},\{0,4\}}(x)+\HA_{\{0,4\},4,\{0,4\}}(x)
\]
\end{example}

\subsection{Mellin representations}

An integral representation of a given nested sum in terms of the Mellin transform \eqref{eq:Mellintransform} usually does not take the form of just one term $\M{f(x)}{n}$. In general, Mellin representations of nested sums take the form
\begin{equation}\label{eq:MellinRepresentation}
 c_0+\sum_{j=1}^kc_j^n\M{f_j(x)}{n},
\end{equation}
for some $k\ge1$ with $c_0,\dots,c_k$ and $f_1(x),\dots,f_k(x)$ being independent of $n$. Often, the integral \eqref{eq:Mellintransform} defining the Mellin transform needs to be regularized accordingly due to singularities of $f_j(x)$. Such Mellin representations can be used to compute asymptotic expansions of complicated expressions involving nested sums as described by Eqs. (2.15) and (2.16) from Ref.~\cite{JMP2014}, like it was done e.g.\ in Ref.~\cite{NPB2014}.

In Ref.~\cite{InvMellin}, a general method for computing Mellin representations of P-finite sequences based on constructing and solving differential equations is presented. Specialized on sequences that are given as nested sums, a refined version \cite{InvMellinRefined} of that method was given later, which exploits the nested structure of the input but still relies on constructing and solving differential equations.

Instead, one can use basic identities of the Mellin transform to exploit the structure of the nested sum to be represented. Several basic identities that allow to build Mellin representations of sequences from Mellin representations of simpler sequences are collected in Sec.~2 of Ref.~\cite{JMP2014}, for example. Among them,
\begin{equation}\label{eq:MellinSum}
 \sum_{i=1}^nc^i\M{f(x)}{i}=c^n\M{\frac{x}{x-\frac{1}{c}}f(x)}{n}-\M{\frac{x}{x-\frac{1}{c}}f(x)}{0},
\end{equation}
which reduces the Mellin representation of a sum to the Mellin representation of its summand, as well as
\begin{equation}\label{eq:MellinProduct}
 \M{f(x)}{n}{\cdot}\M{g(x)}{n}=\M{\int_x^1\frac{f(\frac{x}{t})g(t)}{t}\,dt}{n},
\end{equation}
which allows to compute the Mellin representation of a product from the Mellin representations of the factors by evaluating Mellin convolution integrals.

We see that performing a Mellin convolution amounts to computing a definite integral depending on a parameter. An overview of this topic and related algorithms is given in Ref.~\cite{RaabTut}, for example. Here, we just give a short explanation. There are essentially two main approaches in computer algebra for treating rather general parameter integrals, which we briefly compare below without going into details. The key concept is that of \emph{creative telescoping}, which, for a given integrand $f(t)$ depending on additional parameters, aims to construct a linear operator $L$ such that $L$ commutes with $\frac{d}{dt}$ and an explicit antiderivative $g(t)$ of $L(f(t))$ can be found:
\begin{equation}\label{eq:CT}
 L(f(t)) = \frac{d}{dt}g(t).
\end{equation}
Then, by the properties imposed on $L$, integrating from $a$ to $b$ yields an implicit equation $L\left(\int_a^bf(t)\,dt\right) = g(b)-g(a)$ for the parameter integral, which typically is a differential or recurrence equation depending on how $L$ acts on the parameters in the integrand. If $a$ or $b$ depends on additional parameters acted on by $L$, then additional terms arise in the equation from the difference $L\left(\int_a^bf(t)\,dt\right)-\int_a^bL(f(t))\,dt$.
To obtain an evaluation of the parameter integral, the implicit equation still has to be solved by other means.
For computing telescoping relations \eqref{eq:CT}, on the one hand, many algorithms based on \emph{holonomic systems} have been developed over the past 30 years, see e.g.\ \cite{Zeilberger,Chyzak,ChenKauersKoutschan,ChenHoeijKauersKoutschan,BostanChyzakLairezSalvy,Hoeven}. 
On the other hand, integration algorithms based on \emph{differential fields} have been developed for more than 50 years, many of which are suitable for creative telescoping, see e.g.\ \cite{Risch,CMack,Norman,SingerEtAl,Boettner,Raab}.

One main difference between these two approaches lies in the way functions are represented. Therefore, there is a fundamental difference in what kind of antiderivatives $g(t)$ can be found by the algorithms to fulfill Eq.~\eqref{eq:CT}. In the holonomic systems approach, $g$ is restricted to the form $g(t)=Q(f(t))$ for some linear (typically differential/recurrence) operator $Q$ acting on the integrand. Hence, essentially no additional functions can appear in $g$ that did not already appear in the integrand $f$. In contrast, with algorithms using differential fields, antiderivatives $g(t)$ can be found that involve certain new functions that do not already occur in the integrand $f(t)$. As a result, potentially simpler operators $L$ may allow the antidifferentiation in Eq.~\eqref{eq:CT} to be carried out by the algorithm in question, yielding a differential or recurrence equation of smaller order for the parameter integral.
Another difference lies in the type of integrands that can be handled. In the holonomic systems approach, most general algorithms work with \emph{D-finite functions}, i.e.\ integrands that satisfy linear homogeneous differential equations with polynomial coefficients. Algorithms using differential fields can deal with Liouvillian integrands \cite{SingerEtAl} and also a large class of non-Liouvillian functions, see e.g.\ \cite{Boettner,Raab}. The two classes of integrands accessible by algorithms of the two approaches are very large, each covering a majority of common special functions. Despite the fact that many functions (e.g.\ all d'Alembertian functions) are both D-finite and Liouvillian, there are many Liouvillian functions that are not D-finite, and vice versa. In particular, arbitrary quotients and compositions of Liouvillian functions are Liouvillian again, whereas the same is not true for D-finite functions.
The holonomic systems approach is not specific to integration and many algorithms using this approach can be used for summation as well. Inspired by integration algorithms using differential fields, an algorithmic analog has been introduced for summation by Karr \cite{Karr} using \emph{difference fields}, which was developed further by Schneider, see \cite{SchneiderTheory} and references therein. 

In fact, there is also a third approach to compute parameter integrals, which is more specialized. It relies on collections of identities that either evaluate the given parameter integral or relate it to other integrals. Such identities can be used as rewrite rules for evaluating or simplifying parameter integrals of specific form. Coming back to Mellin convolutions \eqref{eq:MellinProduct}, many such identities are provided in Sec.~4 of Ref.~\cite{JMP2014} specially for rewriting integrals $\int_x^1t^{-1}g(\frac{x}{t})h(t)f(t)\,dt$, with certain concrete choices of $g$ and $h$ but arbitrary $f$, in terms of similar integrals which only involve the derivative $f^\prime$ instead of $f$. If these rewrite rules are applied to $f$ being a nested integral, then in the resulting integral the function $f^\prime$ will only involve a nested integral with lower depth. Iterating this reduction, a Mellin convolution involving a nested integral can be performed, provided sufficiently many rewrite rules are available. This is analogous to rewrite rules like Eqs. \eqref{eq:rule705} through \eqref{eq:rule708} mentioned above, which remove one summation step each.
Due to the page limit, we refer to Ref.~\cite{JMP2014} for examples of such rules an how they can be combined to obtain general patterns for Mellin representations of nested sums with certain prefactors.

There are methods to design new rewrite rules of this type for Mellin convolutions with other choices of $g$ and $h$ or even for other types of parameter integrals, e.g.\ integrals arising from integral transforms. For example, Sec.~7 of Ref.~\cite{JMP2014} also lists some rewrite rules for the integral transform $\int_0^1\frac{1}{1-tx}f(t)\,dt$, that allow to convert Mellin representations of sequences into their generating function. Specialized identities can also be derived for doing other tasks by rewriting, e.g.\ evaluating Mellin transforms in terms of nested sums.

\section{Rationalizing transformations}
\label{sec:transformations}

In this section, we discuss simplification of square roots in the integrands by suitable changes of variables. That is, if $\sqrt{f(x)}$ appears with $f(x)$ a rational function, we want to apply a transformation $x=g(y)$ such that $\sqrt{f(g(y))}$ can be simplified to a rational function. To preserve the nested structure of the integrals properly, the same change of variables has to be applied to all integrands of a nested integral. For instance, a nested integral of the form \eqref{eq:NestedIntegral0} would become the nested integral
\begin{multline}
 \int_{g^{-1}(0)}^{g^{-1}(x)}f_1(g(u_1))g^\prime(u_1)\int_{g^{-1}(0)}^{u_1}f_2(g(u_2))g^\prime(u_2)\dots\\
 \dots\int_{g^{-1}(0)}^{u_{k-1}}f_k(g(u_k))g^\prime(u_k)\,du_k\dots du_1.
\end{multline}
Since this transformation should not introduce any new functions in those integrands that are already rational (nor via $g^\prime(y)$), $g(y)$ is required to be a rational function.

\begin{example}\label{ex:530case40real}
Formula (5.30) in \cite{JMP2014} gives an integral representation of the nested sum
\[
 \sum\limits_{i=1}^n\frac{\binom{2i}{i}}{i^2}\sum\limits_{j=1}^i\frac{(-1)^j}{j^2}.
\]
Among others, it involves the nested integral
\[
 \int_x^1\frac{1}{t_1}\int_{t_1}^1\frac{1}{\sqrt{t_2}\sqrt{1+t_2}}\int_{t_2}^1\frac{1}{\sqrt{t_3}\sqrt{1+t_3}}\int_{t_3}^1\frac{1}{\sqrt{t_4}\sqrt{1-t_4}}\,dt_4\,dt_3\,dt_2\,dt_1
\]
of the form \eqref{eq:NestedIntegral1}. In order to obtain integrands that are rational functions, based on Eqs. \eqref{eq:case40Greal} and \eqref{eq:case40Grealinv} below for $a_1=-1$ and $a_2=1$, we can use the change of variables
\[
 x=\frac{2y^2}{y^4+1} \quad\text{respectively}\quad y=\frac{\sqrt{1+x}-\sqrt{1-x}}{\sqrt{2}\sqrt{x}},
\]
which maps the interval $[0,1]$ bijectively to itself to preserve the form \eqref{eq:NestedIntegral1}. Then, by Eqs. \eqref{eq:case40R1real} and \eqref{eq:case40R2real}, we have $\sqrt{x}\sqrt{1 \pm x} = \frac{\sqrt{2}y(1 \pm y^2)}{y^4+1}$ and the differentials occurring in the above nested integral are transformed as follows.
\[
 \frac{dx}{x} = \left(\frac{2}{y}-\frac{4y^3}{y^4+1}\right)dy \quad\quad\quad
 \frac{dx}{\sqrt{x}\sqrt{1 \pm x}} = \frac{2\sqrt{2}(1 \mp y^2)}{y^4+1}dy
\]
To obtain properly nested integrals again, the change of variable has to be applied uniformly to all levels within the nested integral, which in the present case yields
\[
 32\sqrt{2}\int_y^1\left(\frac{1}{u_1}-\frac{2u_1^3}{u_1^4+1}\right)\int_{u_1}^1\frac{1-u_2^2}{u_2^4+1}\int_{u_2}^1\frac{1-u_3^2}{u_3^4+1}\int_{u_3}^1\frac{1+u_4^2}{u_4^4+1}\,du_4\,du_3\,du_2\,du_1
\]
having rational integrands.
Altogether, the nested integral above can hence be written in terms of cyclotomic harmonic polylogarithms \cite{Cyclotomic} evaluated at $\frac{\sqrt{1+x}-\sqrt{1-x}}{\sqrt{2}\sqrt{x}}$.
\end{example}

Below, we let $C$ be any field of characteristic zero, i.e.\ a field extension of the rational numbers possibly containing indeterminates. Its algebraic closure is denoted by $\overline{C}$. Elements of $C$ will be considered as constants.
In full generality, the problem of finding a rationalizing transformation for square roots of univariate rational functions can be stated as follows.

\begin{problem}\label{prob:rationalize}
Given a set of nonzero rational functions $F \subset C(x)$, find, if possible, a non-constant rational function $g \in C(y)$ such that, upon substituting $g(y)$ for $x$, every element of $F$ can be written in the form $c{\cdot}f(y)^2$ for some $c \in C$ and $f \in C(y)$.
\end{problem}

\begin{remark}\label{rem:parameterization}
Not for every set of radicands $F \subset C(x)$ a rationalizing transformation $g \in C(y)$ exists, however. If $F$ is such that there are $f_1,\dots,f_n \in F$ and $r \in C(x)$ such that $p:=f_1{\cdot}\dots{\cdot}f_n{\cdot}r^2$ is a squarefree polynomial (i.e.\ not divisible by the square of any non-constant polynomial) of degree $\ge3$, then a rationalizing transformation cannot exist for $F$. This is because, for any rationalizing transformation $g \in C(y)$, there would exist nonzero $c \in C$ and $f \in C(y)$ such that $p(g(y))=c{\cdot}f(y)^2$. However, the irreducible algebraic curve defined by $p(X)=c{\cdot}Y^2$ has a rational parameterization $X=g(y)$ and $Y=f(y)$ with $f,g \in \overline{C}(y)$ if and only if the curve has genus $0$ (see e.g.\ \cite{RatCurvesBook}), which happens only for $\deg_x(p)\le2$ since $p$ is squarefree.
\end{remark}

In the past, many ad-hoc transformations have been used in practice for explicitly converting integrands with square roots into rational integrands, see e.g. \cite{Substitutions}, or for evaluating certain quantities in terms of polylogarithms at arguments involving square roots, see e.g.\ \cite{BroadhurstFleischerTarasov,FleischerKotikovVeretin}. In the rest of this section, we will give an exhaustive overview of explicit formulae for rationalizing transformations solving Problem~\ref{prob:rationalize}. In general, also a multivariate analog of Problem~\ref{prob:rationalize} can be considered, for which a general method to construct rationalizing transformations was given recently \cite{BesierStratenWeinzierl} that works in many cases and has been implemented \cite{BesierWasserWeinzierl}.

\begin{remark}\label{rem:ReductionOfSquares}
In all that follows, we exclude the case when all elements of $F$ are squares in $C(x)$, which trivially admits the rationalizing transformation $g(y)=y$. Without loss of generality, we can assume that the elements of $F$ are monic squarefree polynomials in $C[x]$, since multiplying an element of $F$ by a nonzero square in $C(x)$ or by a nonzero constant from $C$ does not change the possible transformations $g$. Furthermore, we can assume without loss of generality that no polynomial in $F$ divides another, since distinct $f_1,f_2 \in F$ with $f_1|f_2$ allow to replace $f_2$ by the quotient $f_2/f_1$ (or to remove $f_1$ from $F$, if $f_1$ is constant) without changing $g$.
\end{remark}

For simplicity, one is interested in rationalizing transformations $g$ of low degree $\max(\deg(\num(g)),\deg(\den(g)))$. As soon as one rationalizing transformation $g$ is known for a given set $F$, infinitely many rationalizing transformations can be obtained by composition $g(h(y))$ with any non-constant $h \in C(y)$. If $h \in C(y)$ has degree $1$, i.e.\ it is of the form $\frac{ay+b}{cy+d}$, then composition $g(h(y))$ does not change the degree of $g$, otherwise the degree is increased.

Next, we give explicit formulae for rationalizing transformations for radicands which have coefficients in arbitrary field extensions of $\mathbb{Q}$. The transformations are chosen such that they map $y=0$ to $x=g(0)=0$. After that, we will give dedicated formulae for the fields $C=\mathbb{R}$ and $C=\mathbb{C}$, where we impose in addition that the transformations map the interval $[0,1]$ bijectively to itself. All rationalizing transformations given below have been constructed by the author on various occasions distributed over the past few years, using the computer algebra systems \MMA, \Singular, and \Maple, with verification of their properties in \MMA.

\subsection{General transformations mapping $0$ to $0$}
\label{sec:generaltransformations}

Let $C$ be a field of characteristic zero. Any $F \subset C(x)$, for which a rationalizing transformation exists, see Remark~\ref{rem:parameterization}, can be reduced, as described in Remark~\ref{rem:ReductionOfSquares}, to one of the following four essentially different cases (or to the trivial case $F=\{\}$).

Note that all rationalizing transformations $g \in C(y)$ given below are of lowest possible degree $\max(\deg(\num(g)),\deg(\den(g)))$. Therefore, in each of the cases, any other rationalizing transformation can be obtained via composition $g(h(y))$ with $h \in \overline{C}(y)$. Moreover, the coefficients in the given transformations are rational expressions in the coefficients of the radicand polynomials from $F$. More precisely, if the field $C$ is the smallest field extension of $\mathbb{Q}$ that contains the coefficients of these polynomials, then the coefficients of $g$ lie in the same field $C$ and do not involve any new algebraic or transcendental numbers.

Expressions for an inverse of $g$ are not unique and may require new algebraic numbers in general. All inverses below are expressed as elements of $\overline{C}(x)[r_1,\dots,r_k]$, where $r_1,\dots,r_k$ are square roots of the polynomials in $F$. In addition to directly satisfying $g(g^{-1}(x))=x$, the formulae for $g^{-1}(x)$ below are selected to yield the unique Puiseux series in $\overline{C}((x^{1/2}))$ that also satisfies $g^{-1}(g(y))=y$.

We start with the simplest case of one linear polynomial
\begin{equation}\label{eq:case1F}
 F=\{x-a\},
\end{equation}
where $a \in C$. Then, for instance, we can use one of the following two rationalizing transformations of degree $2$, depending on whether $a$ is zero or not.
\begin{align}
 g(y) &= y^2\label{eq:case10G}\\
 g(y) &= -4ay(y+1)\label{eq:case1G}
\end{align}
An inverse transformation is straightforwardly obtained as
\begin{equation}\label{eq:case1Ginv}
 g^{-1}(x) = \sqrt{x} \quad\text{respectively}\quad g^{-1}(x) = \frac{\sqrt{x-a}-\sqrt{-a}}{2\sqrt{-a}}.
\end{equation}

Next, we consider one quadratic radicand polynomial
\begin{equation}\label{eq:case2F}
 F=\{x^2+c_1x+c_0\},
\end{equation}
where $c_0,c_1 \in C$ are such that $c_1^2 \neq 4c_0$, i.e.\ the polynomial is not a square. Depending on whether $c_0$ is zero or not, we can use one of the following two rationalizing transformations of degree $2$, for example.
\begin{align}
 g(y) &= \frac{c_1y^2}{4(y+1)}\label{eq:case20G}\\
 g(y) &= \frac{4c_0y}{(c_1^2-4c_0)y^2-2c_1y+1}\label{eq:case2G}
\end{align}
Also in this case, the respective algebraic inverses are straightforwardly obtained.
\begin{align}
 g^{-1}(x) &= \frac{2}{c_1}\left(x+\sqrt{x^2+c_1x}\right)\label{eq:case20Ginv}\\
 g^{-1}(x) &= \frac{c_1x+2c_0-2\sqrt{c_0}\sqrt{x^2+c_1x+c_0}}{(c_1^2-4c_0)x}\label{eq:case2Ginv}
\end{align}

By Remark~\ref{rem:parameterization}, $F$ cannot contain a squarefree polynomial of degree $\ge3$. So, we continue with the case of two linear polynomials
\begin{equation}\label{eq:case3F}
 F=\{x-a_1,x-a_2\},
\end{equation}
where $a_1,a_2 \in C$ are distinct. Again, we distinguish the two cases whether $0$ is a root of one of these polynomials or not. Assuming $a_1\neq0$, without loss of generality, we can use one of the following two rationalizing transformations of degree $4$, depending on whether $a_2$ is zero or not.
\begin{align}
 g(y) &= \frac{4a_1y^2}{(y^2+1)^2}\label{eq:case30G}\\
 g(y) &= \frac{4a_1a_2y(y-a_1)(y-a_2)}{(y^2-a_1a_2)^2}\label{eq:case3G}
\end{align}
Writing the inverse transformations in terms of square roots of polynomials in $F$, we obtain the following expressions for the inverse of \eqref{eq:case30G} and \eqref{eq:case3G}, respectively.
\begin{align}
 g^{-1}(x) &= \frac{\sqrt{a_1}\sqrt{x}}{\sqrt{-a_1}x}\left(\sqrt{-a_1}-\sqrt{x-a_1}\right)\label{eq:case30Ginv}\\
 g^{-1}(x) &= \frac{\left(a_1+\sqrt{-a_1}\sqrt{x-a_1}\right)\left(a_2+\sqrt{-a_2}\sqrt{x-a_2}\right)}{x}\label{eq:case3Ginv}
\end{align}

Finally, we conclude with the case of two quadratic polynomials, which by Remark~\ref{rem:parameterization} need to have a common root. This is equivalent to the more symmetric case of three quadratic polynomials that pairwise have exactly one common root, i.e.
\begin{equation}\label{eq:case4F}
 F=\{(x-a_1)(x-a_2),(x-a_1)(x-a_3),(x-a_2)(x-a_3)\},
\end{equation}
where $a_1,a_2,a_3 \in C$ are pairwise distinct. Assuming $0 \not\in \{a_1,a_2\}$, without loss of generality, we can use one of the following two rationalizing transformations of degree $4$, for instance, depending on whether $a_3$ is zero or not.
\begin{align}
 g(y) &= \frac{4a_1a_2y^2}{(a_1-a_2)^2y^4+2(a_1+a_2)y^2+1}\label{eq:case40G}\\
 g(y) &= -\frac{4a_1a_2a_3y(y-a_1)(y-a_2)(y-a_3)}{(s_1^2-4s_2)y^4+8s_3y^3-2s_1s_3y^2+s_3^2}\label{eq:case4G}
\end{align}
For shorter notation in Eq.~\eqref{eq:case4G}, we used the elementary symmetric polynomials
\begin{equation}\label{eq:ElSymA}
 s_1=a_1+a_2+a_3,\quad s_2=a_1a_2+a_1a_3+a_2a_3,\quad\text{and}\quad s_3=a_1a_2a_3.
\end{equation}
Using square roots of only two of the three polynomials in $F$, we can write the respective inverses of \eqref{eq:case40G} and \eqref{eq:case4G} as follows. For shorter notation, we abbreviate the square roots $r_1:=\sqrt{(x-a_1)(x-a_2)}$ and $r_2:=\sqrt{(x-a_1)(x-a_3)}$ in Eq.~\eqref{eq:case4Ginv}.
\begin{align}
 g^{-1}(x) &= \frac{\sqrt{a_1a_2}}{(a_1-a_2)x}\left(\frac{\sqrt{x(x-a_1)}}{\sqrt{-a_1}}-\frac{\sqrt{x(x-a_2)}}{\sqrt{-a_2}}\right)\label{eq:case40Ginv}\\
 g^{-1}(x) &= \frac{s_3}{(s_1^2-4s_2)x+4s_3}\Bigg(s_1-2x-\frac{s_1-2a_3}{\sqrt{a_1a_2}}r_1\nonumber\\
 &\quad-\frac{s_1-2a_2}{\sqrt{a_1a_3}}r_2+\frac{a_1{\cdot}(s_1-2a_1)}{\sqrt{a_1a_2}\sqrt{a_1a_3}(x-a_1)}r_1r_2\Bigg)\label{eq:case4Ginv}
\end{align}

\subsection{Real-valued square roots on the interval $[0,1]$}
\label{sec:realvalued}

Here, we consider $C=\mathbb{R}$ and assume that all radicands $f \in F$ are such that $f(x)\ge0$ for all $x \in [0,1]$, so that all the square roots are real-valued on $[0,1]$. In addition, we require rationalizing transformations which not only map $0$ to $0$ like the ones above but which also map the interval $[0,1]$ bijectively to itself and hence are monotonically increasing on that interval. Such bijections of $[0,1]$ allow to preserve this common integration range of integrals and avoid non-real integration bounds. Indeed, as the exhaustive collection below shows, this is possible for all cases discussed above, where rationalizing transformations exist at all, and the degrees of the transformations remain the same. This is achieved by carefully constructed M\"{o}bius transformations of $y$ in the general formulae given in Section~\ref{sec:generaltransformations}.

For each rationalizing transformation $g(y)$, we give an explicit formula for the inverse $g^{-1}(x)$ so that, for all $x \in [0,1]$, the unique $y \in [0,1]$ with $x=g(y)$ is given by $y=g^{-1}(x)$.
As above, it will be a common property of all rationalizing transformations given below that their inverse is given in terms of the square roots of the original radicand polynomials. In order to avoid additional case distinctions, we use radicand polynomials like $a^{-1}(a-x)$ instead of $x-a$ and $a-x$, for instance. It is straightforward to adapt the formulae to other normalizations of radicands if necessary.
Outside the interval $[0,1]$, unless stated otherwise, the explicit expressions given below for $g(y)$ and $g^{-1}(x)$ still satisfy $g(g^{-1}(x))=x$, but $g^{-1}(g(y))=y$ need not hold due to the multivalued nature of the inverse of $g$.

Once a rationalizing transformation $g(y)$ is known for a given set $F$ that maps $[0,1]$ to itself, infinitely many such rationalizing transformations of the same degree can be obtained by composition $g(h(y))$ with
\begin{equation}\label{eq:IntroduceLambda}
 h(y) = \frac{y}{(1-\lambda)y+\lambda},
\end{equation}
where $\lambda>0$ is arbitrary. In fact, all M\"{o}bius transformations that map $[0,1]$ to itself in a bijective and monotonically increasing way are given by Eq.~\eqref{eq:IntroduceLambda} for some $\lambda>0$. The special value $\lambda=1$ yields the identity map and replacing $\lambda$ by $1/\lambda$ gives the inverse transformation of \eqref{eq:IntroduceLambda}.

In contrast to Section~\ref{sec:generaltransformations} above, however, it is no longer possible in all cases to give transformations of the same degree which have coefficients that are rational expressions in the coefficients of the radicand polynomials from $F$. More precisely, the coefficients of the transformations given below involve taking square roots, but still they are real numbers. Hence, if we take $C$ as the smallest field extension of $\mathbb{Q}$ that contains the coefficients of the polynomials in $F$, then the coefficients of the transformations may lie in an algebraic extension of $C$. The rationalizing transformations below are chosen so that the degree of this field extension over $C$ is minimal.

\subsubsection{One square root}

Starting with the simplest case of just a linear radicand, the radicand polynomial (up to a positive constant factor) has to have the form
\begin{equation}\label{eq:case1Freal}
 x \quad\text{or}\quad a^{-1}(a-x),
\end{equation}
for some $a<0$ or $a\ge1$, in order to be non-negative for all $x \in [0,1]$. In the former case, the rationalizing transformation \eqref{eq:case10G} already has the property of mapping the interval $[0,1]$ bijectively to itself, so we trivially have
\begin{equation}\label{eq:case10Greal}
 g(y) = y^2 \quad\text{and}\quad g^{-1}(x) = \sqrt{x}
\end{equation}
on $[0,1]$. In the latter case, however, we need to modify the transformation \eqref{eq:case1G}. Introducing $\alpha := \sqrt{1-a^{-1}} \ge 0$ for any $a<0$ or $a\ge1$, we have that both
\begin{align}
 g(y) &= y{\cdot}\big((1-2a+2a\alpha)y+2a{\cdot}(1-\alpha)\big)\label{eq:case1Greal} \quad\text{and}\\
 g^{-1}(x) &= a{\cdot}(1+\alpha)\left(1-\sqrt{a^{-1}(a-x)}\right)\label{eq:case1Grealinv}
\end{align}
map the interval $[0,1]$ bijectively to itself in a monotonically increasing way.
The M\"{o}bius transformation used to obtain \eqref{eq:case1Greal} from \eqref{eq:case1G} is $h(y)=-\frac{y}{2a(1+\alpha)}$.
From Eq.~\eqref{eq:case10Greal} resp.\ \eqref{eq:case1Grealinv}, we easily obtain also the explicit expressions of the square root
\begin{equation}\label{eq:case1Rreal}
 \sqrt{x}=y \quad\text{respectively}\quad \sqrt{a^{-1}(a-x)}=(\alpha-1)y+1
\end{equation}
as rational functions in $y$ whenever $y=g^{-1}(x)$.
For later reference, we note that the inverses $g^{-1}(x)$ given by Eqs. \eqref{eq:case10Greal} and \eqref{eq:case1Grealinv} can be rewritten in the form
\begin{equation}\label{eq:case1Grealinv2}
 g^{-1}(x) = \frac{x}{\sqrt{x}} \quad\text{respectively}\quad g^{-1}(x) = \frac{(1+\alpha)x}{1+\sqrt{a^{-1}(a-x)}}.
\end{equation}

Next, we consider radicands that are quadratic polynomials as in Eq.~\eqref{eq:case2F}. First, we assume that $x=0$ is a root of the radicand, which then, being squarefree and nonnegative on the interval $[0,1]$, necessarily (up to a positive constant factor) equals
\begin{equation}\label{eq:case20Freal}
 a^{-1}x(a-x),
\end{equation}
for some $a<0$ or $a\ge1$.
For $a<0$ or $a>1$, the transformations
\begin{align}
 g(y) &= \frac{ay^2}{y^2+a-1}\label{eq:case20Greal} \quad\text{and}\\
 g^{-1}(x) &= a\alpha\frac{\sqrt{a^{-1}x(a-x)}}{a-x}\label{eq:case20Grealinv}
\end{align}
map the interval $[0,1]$ bijectively to itself in a monotonically increasing way, where the inverse transformation \eqref{eq:case20Grealinv} involves $\alpha := \sqrt{1-a^{-1}} > 0$. In fact, Eq.~\eqref{eq:case20Greal} can be obtained from Eq.~\eqref{eq:case20G} with $c_1=-a$ by substituting $h(y)=-\frac{2y}{y\pm\sqrt{1-a}}$ for $y$, which has complex coefficients if $a>1$ even though the transformations \eqref{eq:case20G} and \eqref{eq:case20Greal} have real coefficients.
Note that Eqs. \eqref{eq:case20Greal} and \eqref{eq:case20Grealinv} are not valid if $a=1$. For the special value $a=1$, the following rationalizing transformation maps the interval $[0,1]$ bijectively to itself in a monotonically increasing way, for example.
\begin{align}
 g(y) &= \frac{y^2}{2y^2-2y+1}\label{eq:case201Greal}\\
 g^{-1}(x) &= \frac{x-\sqrt{x(1-x)}}{2x-1}\label{eq:case201Grealinv}
\end{align}
Observe that the singularity at $x=\frac{1}{2}$ in Eq.~\eqref{eq:case201Grealinv} is removable. The transformation \eqref{eq:case201Greal} can be obtained from Eq.~\eqref{eq:case20G} with $c_1=-1$ by replacing $y$ with $h(y)=-\frac{2y}{(1-i)y+i}$ or its complex conjugate.
By Eq.~\eqref{eq:case20Grealinv} respectively \eqref{eq:case201Grealinv}, the square root is easily expressed as rational function in $y=g^{-1}(x)$ by
\begin{equation}\label{eq:case20Rreal}
 \sqrt{a^{-1}x(a-x)} = \frac{a\alpha y}{y^2+a-1} \quad\text{respectively}\quad \sqrt{x(1-x)} = \frac{y(1-y)}{2y^2-2y+1}.
\end{equation}
Furthermore, the inverses given by Eqs. \eqref{eq:case20Grealinv} and \eqref{eq:case201Grealinv} can also be written as
\begin{equation}\label{eq:case20Grealinv2}
 g^{-1}(x) = \frac{\alpha x}{\sqrt{a^{-1}x(a-x)}} \quad\text{respectively}\quad g^{-1}(x) = \frac{x}{x+\sqrt{x(1-x)}}.
\end{equation}

If $x=0$ is not a root of the quadratic radicand, then (up to a positive constant factor) the quadratic polynomial has the form
\begin{equation}\label{eq:case2Freal}
 c_0^{-1}(x^2+c_1x+c_0),
\end{equation}
where $c_0,c_1 \in \mathbb{R}$, $c_0 \neq 0$, are such that the polynomial is not a square and is nonnegative for all $x \in [0,1]$. These conditions are equivalent to requiring $c_0,c_1 \in \mathbb{R}$ to be such that the conditions $c_0 \neq 0$, $c_1^2 \neq 4c_0$, and $c_0(c_0+c_1+1)\ge0$ hold and at least one of the inequalities $-1<2c_0+c_1<0$ and $c_1^2>4c_0$ does not hold. In this case, we set $\alpha := \sqrt{1+c_0^{-1}{\cdot}(c_1+1)} \ge 0$ and the following rationalizing transformation maps the interval $[0,1]$ bijectively to itself in a monotonically increasing way.
\begin{align}
 g(y) &= \frac{y(c_1y+2c_0{\cdot}(1+\alpha))}{1+2c_0+c_1+2c_0\alpha-y^2}\label{eq:case2Greal}\\
 g^{-1}(x) &= c_0{\cdot}(1+\alpha)\frac{\sqrt{c_0^{-1}(x^2+c_1x+c_0)}-1}{x+c_1}\label{eq:case2Grealinv}
\end{align}
In Eq.~\eqref{eq:case2Grealinv}, the singularity at $x=-c_1$ is removable. Note that these formulae work regardless whether the quadratic polynomial \eqref{eq:case2Freal} has two real roots (i.e.\ $c_1^2>4c_0$) or two conjugate complex roots (i.e.\ $c_1^2<4c_0$). We can obtain Eq.~\eqref{eq:case2Greal} also from Eq.~\eqref{eq:case2G} via the M\"{o}bius transformation $h(y)=\frac{y}{c_1y+2c_0(1+\alpha)}$. By virtue of Eq.~\eqref{eq:case2Grealinv}, we have the following expression of the square root as rational function in terms of $y=g^{-1}(x)$.
\begin{equation}\label{eq:case2Rreal}
 \sqrt{c_0^{-1}(x^2+c_1x+c_0)} = \frac{1+2c_0+c_1+2c_0\alpha+c_1(1+\alpha)y+y^2}{1+2c_0+c_1+2c_0\alpha-y^2}
\end{equation}
In fact, the inverse \eqref{eq:case2Grealinv} can also be written in different form.
\begin{equation}\label{eq:case2Grealinv2}
 g^{-1}(x) = \frac{(1+\alpha)x}{1+\sqrt{c_0^{-1}(x^2+c_1x+c_0)}}
\end{equation}

\subsubsection{Two square roots}

For two real-valued square roots of linear polynomials on the interval $[0,1]$, up to a positive constant factor, the two radicand polynomials are given by the set
\begin{equation}\label{eq:case30Freal}
 F=\{x,a^{-1}(a-x)\}
\end{equation}
for some $a<0$ or $a\ge1$, if one of the square roots vanishes at $x=0$. The generic case when neither of the two square roots vanishes at $x=0$ gives rise to radicands \eqref{eq:case3Freal} below. First, for the radicands \eqref{eq:case30Freal}, we introduce $\alpha:=\sqrt{1-a^{-1}}\ge0$ to express the rationalizing transformation
\begin{align}
 g(y) &= \frac{4y^2}{((1-\alpha)y^2+1+\alpha)^2}\label{eq:case30Greal}\\
 g^{-1}(x) &= a{\cdot}(1+\alpha)\frac{1-\sqrt{a^{-1}(a-x)}}{\sqrt{x}}\label{eq:case30Grealinv}
\end{align}
mapping the interval $[0,1]$ bijectively to itself in a monotonically increasing way. The limit of formula \eqref{eq:case30Grealinv} at $x=0$ is finite and zero.
Although, with $a_1=a$, both transformations \eqref{eq:case30G} and \eqref{eq:case30Greal} have real coefficients, the M\"{o}bius transformation $h(y)=\frac{y}{\sqrt{2a-1+2a\alpha}}$, which changes the former into the latter, has complex coefficients if $a<0$.
Based on Eqs. \eqref{eq:case30Greal} and \eqref{eq:case30Grealinv}, the rational function representations
\begin{equation}\label{eq:case30Rreal}
 \sqrt{x}=\frac{2y}{(1-\alpha)y^2+1+\alpha} \quad\text{and}\quad \sqrt{a^{-1}(a-x)}=\frac{1-2a(1+\alpha)+y^2}{1-2a(1+\alpha)-y^2}
\end{equation}
of the square roots in terms of $y=g^{-1}(x)$ hold.
Also in this case, the inverse \eqref{eq:case30Grealinv} can be rewritten to obtain the following expression.
\begin{equation}\label{eq:case30Grealinv2}
 g^{-1}(x) = \frac{(1+\alpha)\sqrt{x}}{1+\sqrt{a^{-1}(a-x)}}
\end{equation}

If none of the two square roots vanishes at $x=0$, the radicand polynomials (up to a positive constant factor) are given by
\begin{equation}\label{eq:case3Freal}
 F=\{a_1^{-1}(a_1-x),a_2^{-1}(a_2-x)\},
\end{equation}
with distinct $a_1,a_2 \in \mathbb{R}$ such that $a_i<0$ or $a_i\ge1$ for each $i$. With
\begin{equation}\label{eq:case3Alphareal}
 \alpha:=\left(1+\sqrt{1-a_1^{-1}}\right)\left(1+\sqrt{1-a_2^{-1}}\right)>1,
\end{equation}
a rationalizing transformation that maps the interval $[0,1]$ bijectively to itself in a monotonically increasing way can be obtained from Eq.~\eqref{eq:case3G} by replacing $y$ with $\frac{y}{\alpha}$.
\begin{equation}\label{eq:case3Greal}
 g(y) = \frac{4a_1a_2\alpha y(y-a_1\alpha)(y-a_2\alpha)}{(y^2-a_1a_2\alpha^2)^2}
\end{equation}
For $x \in [0,1]$, the inverse can be given as
\begin{equation}\label{eq:case3Grealinv}
 g^{-1}(x) = a_1a_2\alpha\frac{\left(1-\sqrt{a_1^{-1}(a_1-x)}\right)\left(1-\sqrt{a_2^{-1}(a_2-x)}\right)}{x},\end{equation}
where the singularity at $x=0$ is removable. With $y=g^{-1}(x)$ as in Eq.~\eqref{eq:case3Grealinv}, the two square roots become rational functions in $y$ as follows.
\begin{align}
 \sqrt{a_1^{-1}(a_1-x)} &= \frac{y^2-2a_2\alpha y+a_1a_2\alpha^2}{-y^2+a_1a_2\alpha^2}\label{eq:case3R1real}\\
 \sqrt{a_2^{-1}(a_2-x)} &= \frac{y^2-2a_1\alpha y+a_1a_2\alpha^2}{-y^2+a_1a_2\alpha^2}\label{eq:case3R2real}
\end{align}
Equivalently, the inverse \eqref{eq:case3Grealinv} can be written as
\begin{equation}\label{eq:case3Grealinv2}
 g^{-1}(x) = \frac{\alpha x}{\left(1+\sqrt{a_1^{-1}(a_1-x)}\right)\left(1+\sqrt{a_2^{-1}(a_2-x)}\right)}.\end{equation}

\subsubsection{Three square roots}

Dealing with real-valued square roots of three quadratic polynomials, which pairwise have exactly one common root, we first consider the case when $x=0$ is among their roots. Up to a positive constant factor, the three radicand polynomials being nonnegative on the whole interval $[0,1]$ necessarily are of the form
\begin{equation}\label{eq:case40Freal}
 F=\{a_1^{-1}x(a_1-x),a_2^{-1}x(a_2-x),a_1^{-1}a_2^{-1}(a_1-x)(a_2-x)\}
\end{equation}
with distinct $a_1,a_2 \in \mathbb{R}$ such that $a_i<0$ or $a_i\ge1$ for each $i$. Then, a rationalizing transformation and its inverse, both mapping $[0,1]$ bijectively to itself in a monotonically increasing way, can be given as follows.
\begin{align}
 g(y) &= \frac{4s_2y^2}{(-s_1+2s_2(1-\alpha))y^4+2s_1y^2-s_1+2s_2(1+\alpha)}\label{eq:case40Greal}\\
 g^{-1}(x) &= a_1a_2\left(\!\sqrt{1-a_1^{-1}}+\!\sqrt{1-a_2^{-1}}\right)\frac{\!\sqrt{a_1^{-1}x(a_1-x)}-\!\sqrt{a_2^{-1}x(a_2-x)}}{(a_1-a_2)x}\label{eq:case40Grealinv}
\end{align}
For shorter notation in Eq.~\eqref{eq:case40Greal} and also below, we use
\begin{equation}\label{eq:case40Alphareal}
 \alpha:=\sqrt{(1-a_1^{-1})(1-a_2^{-1})}\ge0
\end{equation}
as well as the elementary symmetric polynomials
\begin{equation}\label{eq:ElSymA2}
 s_1=a_1+a_2 \quad\text{and}\quad s_2=a_1a_2
\end{equation}
as abbreviations. Note that the limit of the formula \eqref{eq:case40Grealinv} at $x=0$ is finite and zero. The transformation \eqref{eq:case40Greal} can also be obtained from Eq.~\eqref{eq:case40G} via the M\"{o}bius transformation $h(y)=\frac{y}{\sqrt{-a_1-a_2+2a_1a_2(1+\alpha)}}$, which has complex coefficients if $a_1a_2<0$ even though Eqs. \eqref{eq:case40G} and \eqref{eq:case40Greal} have real coefficients. With $y=g^{-1}(x)$ given by Eq.~\eqref{eq:case40Grealinv}, for $x \in [0,1]$, the three square roots of the polynomials in $F$ can be written as the rational functions
\begin{align}
 \sqrt{a_1^{-1}x(a_1-x)} &= \frac{2s_2y\left((\beta_1-\beta_2)y^2+\beta_1+\beta_2\right)}{(-s_1+2s_2(1-\alpha))y^4+2s_1y^2-s_1+2s_2(1+\alpha)}\label{eq:case40R1real}\\
 \sqrt{a_2^{-1}x(a_2-x)} &= \frac{2s_2y\left((\beta_2-\beta_1)y^2+\beta_1+\beta_2\right)}{(-s_1+2s_2(1-\alpha))y^4+2s_1y^2-s_1+2s_2(1+\alpha)}\label{eq:case40R2real}\\
 \!\sqrt{\frac{(a_1-x)(a_2-x)}{a_1a_2}} &= \frac{(s_1-2s_2(1-\alpha))y^4-s_1+2s_2(1+\alpha)}{(-s_1+2s_2(1-\alpha))y^4+2s_1y^2-s_1+2s_2(1+\alpha)}\label{eq:case40R3real}
\end{align}
in $y$ using also $\beta_i:=\sqrt{1-a_i^{-1}}\ge0$ with $i=1,2$ for shorter notation.
Moreover, the inverse \eqref{eq:case40Grealinv} can be written more symmetrically as
\begin{equation}\label{eq:case40Grealinv2}
 g^{-1}(x) = \left(\!\sqrt{1-a_1^{-1}}+\!\sqrt{1-a_2^{-1}}\right)\frac{x}{\!\sqrt{a_1^{-1}x(a_1-x)}+\!\sqrt{a_2^{-1}x(a_2-x)}}.
\end{equation}
While Eqs. \eqref{eq:case40R1real} and \eqref{eq:case40R2real} also hold for $x$ outside the interval $[0,1]$, Eq.~\eqref{eq:case40R3real} does not hold in the same generality. This restriction can be lifted by replacing the square roots of the polynomials from $F$ by pairwise products of $\sqrt{x}$, $\sqrt{a_1^{-1}(a_1-x)}$, and $\sqrt{a_2^{-1}(a_2-x)}$ in Eqs. \eqref{eq:case40Grealinv} through \eqref{eq:case40Grealinv2}. This modification was used in Example~\ref{ex:530case40real}.

If $x=0$ is not among the roots of the three quadratic polynomials, then (up to a positive constant factor) they are given by
\begin{equation}\label{eq:case4Freal}
 F=\left\{\frac{(a_1-x)(a_2-x)}{a_1a_2},\frac{(a_1-x)(a_3-x)}{a_1a_3},\frac{(a_2-x)(a_3-x)}{a_2a_3}\right\},
\end{equation}
with pairwise distinct $a_1,a_2,a_3 \in \mathbb{R}$ satisfying $a_i<0$ or $a_i\ge1$ for each $i$, since the radicand polynomials are nonnegative for all $x \in [0,1]$. To express the following more compactly, we use the elementary symmetric polynomials \eqref{eq:ElSymA} and we let
\begin{equation}\label{eq:case4Alphareal}
 \alpha:=\!\sqrt{(1-a_1^{-1})(1-a_2^{-1})}+\!\sqrt{(1-a_1^{-1})(1-a_3^{-1})}+\!\sqrt{(1-a_2^{-1})(1-a_3^{-1})}>0.
\end{equation}
Then, replacing $y$ with $\frac{y}{1+\alpha}$ in Eq.~\eqref{eq:case4G}, we obtain the rationalizing transformation
\begin{equation}\label{eq:case4Greal}
 g(y) = -\frac{4s_3y(y-a_1(1+\alpha))(y-a_2(1+\alpha))(y-a_3(1+\alpha))}{(s_1^2-4s_2)y^4+8s_3(1+\alpha)y^3-2s_1s_3(1+\alpha)^2y^2+s_3^2(1+\alpha)^4},
\end{equation}
which maps the interval $[0,1]$ bijectively to itself in a monotonically increasing way. Its inverse on the interval $[0,1]$ is given by
\begin{multline}\label{eq:case4Grealinv}
 g^{-1}(x) = -\frac{s_3(1+\alpha)}{(s_1^2-4s_2)x+4s_3}\left(2x-s_1+(s_1-2a_3)\sqrt{\frac{(a_1-x)(a_2-x)}{a_1a_2}}\right.\\
 \left.+(s_1-2a_2)\sqrt{\frac{(a_1-x)(a_3-x)}{a_1a_3}}+(s_1-2a_1)\sqrt{\frac{(a_2-x)(a_3-x)}{a_2a_3}}\right)
\end{multline}
and, if $s_1^2\neq4s_2$, has a singularity at $x=-\frac{4s_3}{s_1^2-4s_2}\neq0$, which is removable whenever it lies in the interior of the interval $[0,1]$, or at least has a finite limit if $x=1$. On the interval $[0,1]$, the three square roots can be expressed as the rational functions
\begin{equation}\label{eq:case4Rreal}
 \sqrt{\frac{(a_i-x)(a_j-x)}{a_ia_j}} = \frac{p_i(y)p_j(y)}{q(y)}
\end{equation}
in $y$, for $i,j\in\{1,2,3\}$, with $y=g^{-1}(x)$ given by Eq.~\eqref{eq:case4Grealinv}, where for shorter notation the following abbreviations were used.
\begin{align}
 p_i(y) &:= (s_1-2a_i)y^2-2\frac{s_3}{a_i}(1+\alpha)y+s_3(1+\alpha)^2\label{eq:case4Pabbrevreal}\\
 q(y) &:= (s_1^2-4s_2)y^4+8s_3(1+\alpha)y^3-2s_1s_3(1+\alpha)^2y^2+s_3^2(1+\alpha)^4\label{eq:case4Qabbrevreal}
\end{align}
Alternatively, Eq.~\eqref{eq:case4Grealinv} can be written more compactly as
\begin{equation}\label{eq:case4Grealinv2}
 g^{-1}(x) = \frac{(1+\alpha)x}{1+\sqrt{\frac{(a_1-x)(a_2-x)}{a_1a_2}}+\sqrt{\frac{(a_1-x)(a_3-x)}{a_1a_3}}+\sqrt{\frac{(a_2-x)(a_3-x)}{a_2a_3}}}.
\end{equation}
With the expressions \eqref{eq:case4Grealinv} and \eqref{eq:case4Grealinv2} for $g^{-1}(x)$ on the interval $[0,1]$, we do not have $g(g^{-1}(x))=x$ for $x$ outside the interval $[0,1]$ in general (unless one of the $a_i$ is the sum of the other two). To satisfy $g(g^{-1}(x))=x$ as well as Eq.~\eqref{eq:case4Rreal} in full generality, one can replace the square roots of the polynomials from $F$ by the pairwise products of the square roots $\sqrt{a_i^{-1}(a_i-x)}$, $i=1,2,3$, in the formulae \eqref{eq:case4Grealinv}, \eqref{eq:case4Rreal}, and \eqref{eq:case4Grealinv2}.

\begin{remark}
Note the similarity of the formulae \eqref{eq:case1Grealinv2}, \eqref{eq:case20Grealinv2}, \eqref{eq:case2Grealinv2}, \eqref{eq:case30Grealinv2}, \eqref{eq:case3Grealinv2}, \eqref{eq:case40Grealinv2}, and \eqref{eq:case4Grealinv2}. Despite the large variety of rationalizing transformations $g(y)$ given on $[0,1]$, in each case, the inverse can be expressed in the form $g^{-1}(x) = \frac{w(1)x}{w(x)}$, where $w(x)$ is some simple expression in terms of the respective square roots.
\end{remark}

\subsection{Complex-valued square roots on the interval $[0,1]$}
\label{sec:complexvalued}

In the following, we no longer require the square roots to take real values on the interval $[0,1]$ like before in Section~\ref{sec:realvalued}. Moreover, we consider $C=\mathbb{C}$, i.e.\ we also treat radicands with complex coefficients. Still, we aim at rationalizing transformations that, when considered on the interval $[0,1]$, give monotonically increasing bijections of $[0,1]$ to itself. This imposes some restrictions on the rationalizing transformation $g$ and hence also on the radicands in $F$, which we explain now.

If, for a given set $F$ of radicands, there is a rationalizing transformation $g \in \mathbb{C}(y)$ that maps the interval $[0,1]$ to itself, then $g$ necessarily can also be written with real coefficients, since it is real-valued on $[0,1]$ and any rational function $g \in \mathbb{C}(y)$ can be written as $g=g_1+ig_2$ for some $g_1,g_2 \in \mathbb{R}(y)$. Consequently, if some $f \in \mathbb{C}(x)$ becomes a square $f(g(y))$ by the change of variable $x=g(y)$, then also its complex conjugate $\overline{f}$ becomes a square in $\mathbb{C}(y)$ by the same transformation $g$. Altogether, we have that a set $F \subset \mathbb{C}(x)$ does not admit a rationalizing transformation that maps $[0,1]$ to itself, if the set $F \cup \overline{F}$ does not. This restricts most of the general cases listed in Section~\ref{sec:generaltransformations} to special choices of the coefficients of radicands.

For instances that can be reduced (cf.\ Remark~\ref{rem:ReductionOfSquares}) to radicands of real-valued square roots on $[0,1]$, we refer to the transformations in Section~\ref{sec:realvalued}. As it turns out, the remaining cases can also be treated by some of the formulae given in that section. The simplest case not reducible to real-valued square roots is given by one square root whose only singularity lies outside the real line. The set $F \cup \overline{F}$ of radicands can be reduced to the form \eqref{eq:case3Freal} with $a_2=\overline{a_1}\neq{a_1}$. Similarly, for a square root with two singularities, where exactly one of them is on the real line, $F \cup \overline{F}$ can be reduced to the form \eqref{eq:case40Freal} or \eqref{eq:case4Freal} with one $a_i$ being the complex conjugate of one of the others. All other cases of complex-valued roots that admit a rationalizing transformation mapping $[0,1]$ to itself can also be reduced to one of these cases detailed below.

\subsubsection{Two square roots}

We consider $F$ as in Eq.~\eqref{eq:case3Freal}, where $a_1,a_2 \in \mathbb{C}\setminus\mathbb{R}$ are such that $a_2=\overline{a_1}$. With $\alpha$ as in Eq.~\eqref{eq:case3Alphareal}, we have that the rationalizing transformation given by Eq.~\eqref{eq:case3Greal} again maps the interval $[0,1]$ bijectively to itself in a monotonically increasing way. Also the formulae \eqref{eq:case3Grealinv} through \eqref{eq:case3R2real} still hold.

Alternatively, we can express all of these formulae also with real coefficients using $\re(a_1)$, $\im(a_1)$, $|a_1|^2$, and $|\frac{a_1-1}{a_1}|$. More explicitly, $\alpha>1$ can be written as
\begin{equation}\label{eq:case3Alphacomplex}
 \alpha = 1+\left|\frac{a_1-1}{a_1}\right|+\sqrt{2\left(1-\frac{\re(a_1)}{|a_1|^2}+\left|\frac{a_1-1}{a_1}\right|\right)}
\end{equation}
and the rationalizing transformation \eqref{eq:case3Greal} and its inverse \eqref{eq:case3Grealinv} read as follows.
\begin{align}
 g(y) &= \frac{4|a_1|^2\alpha y(y^2-2\re(a_1)\alpha y+|a_1|^2\alpha^2)}{(y^2-|a_1|^2\alpha^2)^2}\label{eq:case3Gcomplex}\\
 g^{-1}(x) &= 4|a_1|^2\alpha\frac{\left(1-\sqrt{a_1^{-1}(a_1-x)}\right)\left(1-\sqrt{\overline{a_1^{-1}}(\overline{a_1}-x)}\right)}{x}\label{eq:case3Gcomplexinv}
\end{align}
Moreover, the two square roots can be written explicitly as $r_1(y)\pm ir_2(y)$ in terms of $y=g^{-1}(x)$ with $r_1,r_2 \in \mathbb{R}(y)$. Then, $r_1(y)$ and $r_2(y)$ give the real and imaginary parts as long as $y$ is real, which happens whenever $x$ is real.
\begin{align}
 \sqrt{a_1^{-1}(a_1-x)} &= \frac{y^2-2\re(a_1)\alpha y+|a_1|^2\alpha^2}{-y^2+|a_1|^2\alpha^2}+i\frac{2\im(a_1)\alpha y}{-y^2+|a_1|^2\alpha^2}\\
 \sqrt{\overline{a_1^{-1}}(\overline{a_1}-x)} &= \frac{y^2-2\re(a_1)\alpha y+|a_1|^2\alpha^2}{-y^2+|a_1|^2\alpha^2}-i\frac{2\im(a_1)\alpha y}{-y^2+|a_1|^2\alpha^2}
\end{align}

\subsubsection{Three square roots}

First, we treat the set of radicands $F$ as in Eq.~\eqref{eq:case40Freal} with $a_1,a_2 \in \mathbb{C}\setminus\mathbb{R}$ such that $a_2=\overline{a_1}$. With $\alpha$ as in Eq.~\eqref{eq:case40Alphareal}, the rationalizing transformation given by Eqs. \eqref{eq:case40Greal} and \eqref{eq:ElSymA2} again maps the interval $[0,1]$ bijectively to itself in a monotonically increasing way. Its inverse is given by Eq.~\eqref{eq:case40Grealinv} and also Eqs. \eqref{eq:case40R1real} through \eqref{eq:case40R3real} remain valid to the extent mentioned there.
While the formula \eqref{eq:case40Grealinv} for $y=g^{-1}(x)$ gives a real value whenever $x$ is real, it does not make Eq.~\eqref{eq:case40R3real} true for general $x$. In the following, we consider the inverse
\begin{equation}
 g^{-1}(x) = a_1a_2\left(\!\sqrt{1-a_1^{-1}}+\!\sqrt{1-a_2^{-1}}\right)\frac{\!\sqrt{a_1^{-1}(a_1-x)}-\!\sqrt{a_2^{-1}(a_2-x)}}{(a_1-a_2)\sqrt{x}}
\end{equation}
instead, which gives the same values if $x>0$, but does not give a real value if $x<0$. In return, the right hand sides of Eqs. \eqref{eq:case40R1real} through \eqref{eq:case40R3real} give correct expressions for the pairwise products of $\sqrt{x}$, $\sqrt{a_1^{-1}(a_1-x)}$, and $\sqrt{a_2^{-1}(a_2-x)}$ for general $x$.

We can write these formulae with real coefficients using $\re(a_1)$, $\im(a_1)$, $|a_1|^2$, and $|\frac{a_1-1}{a_1}|$. With
\begin{equation}
 \alpha = \left|\frac{a_1-1}{a_1}\right|>0,
\end{equation}
we introduce the abbreviations
\begin{align}
 \beta &:= \sqrt{\frac{1}{2}\left(1-\frac{\re(a_1)}{|a_1|^2}+\alpha\right)}\\
 q(y) &:= (|a_1|^2-\re(a_1)-|a_1|^2\alpha)y^4+2\re(a_1)y^2+|a_1|^2-\re(a_1)+|a_1|^2\alpha
\end{align}
for shorter notation. Then, we have that
\begin{align}
 g(y) &= \frac{2|a_1|^2y^2}{q(y)}\\
 g^{-1}(x) &= -i\frac{|a_1|^2\beta}{\im(a_1)}{\cdot}\frac{\sqrt{a_1^{-1}(a_1-x)}-\sqrt{\overline{a_1^{-1}}(\overline{a_1}-x)}}{\sqrt{x}}
\end{align}
and with $y=g^{-1}(x)$ we have the following rational expressions in $y$ for the pairwise products of square roots, which reveal the real and imaginary part as long as $y$ is real.
\begin{align}
 \sqrt{x}\sqrt{\frac{a_1-x}{a_1}} &= \frac{(|a_1|^2-\re(a_1)+|a_1|^2\alpha)y}{\beta q(y)}+i\frac{\im(a_1)y^3}{\beta q(y)}\\
 \sqrt{x}\sqrt{\frac{\overline{a_1}-x}{\overline{a_1}}} &= \frac{(|a_1|^2-\re(a_1)+|a_1|^2\alpha)y}{\beta q(y)}-i\frac{\im(a_1)y^3}{\beta q(y)}\\
 \sqrt{\frac{a_1-x}{a_1}}\sqrt{\frac{\overline{a_1}-x}{\overline{a_1}}} &= \frac{-(|a_1|^2-\re(a_1)-|a_1|^2\alpha)y^4+|a_1|^2-\re(a_1)+|a_1|^2\alpha}{q(y)}
\end{align}

Finally, we turn to radicands $F$ as in Eq.~\eqref{eq:case4Freal} with $a_1 \in \mathbb{R}$ such that $a_1<0$ or $a_1\ge1$ and $a_2,a_3 \in \mathbb{C}\setminus\mathbb{R}$ such that $a_3=\overline{a_2}$. With $\alpha$ as in Eq.~\eqref{eq:case4Alphareal}, the rationalizing transformation given by Eqs. \eqref{eq:case4Greal} and \eqref{eq:ElSymA} is a monotonically increasing bijection of the interval $[0,1]$ to itself and its inverse on $[0,1]$ can be given by Eq.~\eqref{eq:case4Grealinv} resp.\ \eqref{eq:case4Grealinv2}. Furthermore, also the expressions \eqref{eq:case4Rreal} for the three square roots remain valid for $y=g^{-1}(x)$ with $x \in [0,1]$. To obtain formulae that hold for general $x$, we instead use the inverse
\begin{multline}
 g^{-1}(x) = -\frac{s_3(1+\alpha)}{(s_1^2-4s_2)x+4s_3}\left(2x-s_1+(s_1-2a_3)\sqrt{\frac{a_1-x}{a_1}}\sqrt{\frac{a_2-x}{a_2}}\right.\\
 \left.+(s_1-2a_2)\sqrt{\frac{a_1-x}{a_1}}\sqrt{\frac{a_3-x}{a_3}}+(s_1-2a_1)\sqrt{\frac{a_2-x}{a_2}}\sqrt{\frac{a_3-x}{a_3}}\right)
\end{multline}
below. In terms of the real quantities $a_1$, $\re(a_2)$, $\im(a_2)$, $|a_2|^2$, and $|\frac{a_2-1}{a_2}|$, we can write
\begin{equation}
 \alpha = \left|\frac{a_2-1}{a_2}\right|+\sqrt{\frac{a_1-1}{a_1}}\sqrt{2\left(1-\frac{\re(a_2)}{|a_2|^2}+\left|\frac{a_2-1}{a_2}\right|\right)}
\end{equation}
and
\begin{multline}\label{eq:case4Qcomplex}
 q(y) = (a_1^2-4a_1\re(a_2)-4\im(a_2)^2)y^4+8a_1|a_2|^2(1+\alpha)y^3\\
 -2a_1|a_2|^2(a_1+2\re(a_2))(1+\alpha)^2y^2+a_1^2|a_2|^4(1+\alpha)^4
\end{multline}
for Eqs. \eqref{eq:case4Alphareal} and \eqref{eq:case4Pabbrevreal}. Then, we have
\begin{equation}
 g(y) = -4a_1|a_2|^2\frac{y(y-a_1(1+\alpha))(y^2-2\re(a_2)(1+\alpha)y+|a_2|^2(1+\alpha)^2)}{q(y)}
\end{equation}
and
\begin{multline}
 g^{-1}(x) = -\frac{a_1|a_2|^2(1+\alpha)}{(a_1^2-4a_1\re(a_2)-4\im(a_2)^2)x+4a_1|a_2|^2}\Bigg(2x-a_1-2\re(a_2)\\
 +a_1\sqrt{\frac{a_1-x}{a_1}}\left(\sqrt{\frac{a_2-x}{a_2}}+\sqrt{\frac{\overline{a_2}-x}{\overline{a_2}}}\right)+(2\re(a_2)-a_1)\sqrt{\frac{a_2-x}{a_2}}\sqrt{\frac{\overline{a_2}-x}{\overline{a_2}}}\\
 \left.+2i\im(a_2)\sqrt{\frac{a_1-x}{a_1}}\left(\sqrt{\frac{a_2-x}{a_2}}-\sqrt{\frac{\overline{a_2}-x}{\overline{a_2}}}\right)\right).
\end{multline}
In terms of $y=g^{-1}(x)$, the pairwise products of square roots can be expressed as
\begin{multline}\label{eq:case4R1complex}
 \sqrt{\frac{a_1-x}{a_1}}\sqrt{\frac{a_2-x}{a_2}} = \frac{a_1(y^2-2\re(a_2)(1+\alpha)y+|a_2|^2(1+\alpha)^2)p(y)}{q(y)}\\
 -i\frac{2\im(a_2)y(y-a_1(1+\alpha))p(y)}{q(y)}
\end{multline}
and
\begin{multline}\label{eq:case4R2complex}
 \sqrt{\frac{a_1-x}{a_1}}\sqrt{\frac{\overline{a_2}-x}{\overline{a_2}}} = \frac{a_1(y^2-2\re(a_2)(1+\alpha)y+|a_2|^2(1+\alpha)^2)p(y)}{q(y)}\\
 +i\frac{2\im(a_2)y(y-a_1(1+\alpha))p(y)}{q(y)}
\end{multline}
as well as
\begin{equation}\label{eq:case4R3complex}
 \sqrt{\frac{a_2-x}{a_2}}\sqrt{\frac{\overline{a_2}-x}{\overline{a_2}}} = \frac{r(y)}{q(y)},
\end{equation}
where for shorter notation, in addition to Eq.~\eqref{eq:case4Qcomplex}, we also use the abbreviations
\begin{equation}
 p(y) := (2\re(a_2)-a_1)y^2-2|a_2|^2(1+\alpha)y+a_1|a_2|^2(1+\alpha)^2
\end{equation}
and
\begin{multline}
 r(y) := (a_1^2+4\im(a_2)^2)y^4-4a_1(a_1\re(a_2)+2\im(a_2)^2)(1+\alpha)y^3\\
 +6a_1^2|a_2|^2(1+\alpha)^2y^2-4a_1^2\re(a_2)|a_2|^2(1+\alpha)^3y+a_1^2|a_2|^4(1+\alpha)^4.
\end{multline}
Also, Eqs. \eqref{eq:case4R1complex} through \eqref{eq:case4R3complex} exhibit the real and imaginary parts whenever the quantity $y=g^{-1}(x)$ is real.

\begin{acknowledgement}
The author wants to thank Jakob Ablinger for useful discussions. The author is supported by the Austrian Science Fund (FWF) grant P~31952.
\end{acknowledgement}

\end{document}